\documentclass[aps,prl,showpacs,
 lengthcheck]{revtex4}
 \usepackage{amssymb,amsmath}
\usepackage{graphicx} 
\graphicspath{{./figures-lettre/}}

\newcommand{\rmd}{{\mathrm{d}}}

\newlength{\bilderlength}

\begin{document}

\title{\sffamily\bfseries\large Systematic Field Theory of the RNA Glass Transition} 

\author{\sffamily\bfseries\normalsize Francois David$^{(1)}$ and Kay J\"org Wiese$^{(2)}$}
\affiliation{$^{(1)}$Service de Physique ThŽorique%
\footnote{URA 2036 of CNRS}, CEA Saclay, 91191 Gif sur Yvette Cedex, France\\
$^{(2)}$Laboratoire de Physique Th\'eorique de l'Ecole Normale
Sup\'erieure, 24 rue Lhomond, 75005 Paris, France}

\date{\small\today}

\begin{abstract}
We prove that the L\"assig-Wiese (LW) field theory for the freezing
transition of the secondary structure of random RNA is renormalizable
to all orders in perturbation theory. The proof relies on a
formulation of the model in terms of random walks and on the use of
the multilocal operator product expansion. Renormalizability allows us
to work in the simpler scheme of open polymers, and to obtain the
critical exponents at 2-loop order. It also allows to prove some exact
exponent identities, conjectured in LW.
\end{abstract}


\maketitle

{T}{ogether} with DNA and proteins, RNA plays a key role in biology.
As such it is important to understand their spatial
conformations. While for proteins the lowest-energy fold depends
strongly on the chemical constitution, and is only tractable
numerically, the problem for RNA is simpler, due to a clear separation
in energy scale between primary structure (the sequence), secondary
structure (pairing of bases in a fold) and tertiary structure
(embedding of a fold in 3-d space). The homopolymer problem (all bases
identical) was solved in 1968 by de Gennes \cite{deGennes}.  He finds
that the pairing-probability of two RNA-bases with labels $s$ and
$t$, counted along the backbone, scales like ${\cal P} (s,t)\sim
|s-t|^{-\rho_{0}}$, with $\rho_{0}=3/2$.  Real RNA molecules however
consist of a sequence of 4 different bases, and their optimal fold
depends on this sequence. Experimentally important (see e.g.\
\cite{Tinoco}) is further the observation, that pairings $(s,t)$ and
$(s',t')$ are either nested ($s<s'<t'<t$) or independent
($s<t<s'<t'$), which graphically amounts to the rule to draw the
sequence and the pairings on the plane without self-intersections
(planarity).
While the problem of a biological RNA-sequence is best solved
numerically, for reference it is crucial to understand the
physics of (planar) pairings of a {\em random} sequence. This was
pioniered by Bundschuh and Hwa (BH) \cite{BH}. They consider a random
pairing model with partition function ($\beta=1/k_BT$)
\begin{equation}
\label{k1} Z_\eta=\sum_{\Phi}\exp\Big[-\beta\sum_{1\le s<t\le
L}\eta(s,t)\Phi(s,t)\Big]
\ ,
\end{equation}
which is defined as a sum over all {\em planar} pairings $\Phi$, such
that $\Phi(s,t)=1$ if $(s,t)$ is a Watson-Crick pair, and $0$
otherwise.  
The pair energy $\eta(s,t)$ is considered as a quenched Gaussian
disorder variable $\eta(s,t)$, with
\begin{equation*}
\label{k2} \overline{\eta(s,t)}=f,\quad
\overline{\eta(s,t)\eta(u,v)}{-}f^2=
\sigma^2\delta{({s{-}u})}\delta{({t{-}v})}.
\end{equation*}
Note that this is an additional approximation from the model of a
random sequence \cite{seq}. A key feature of the above model is that
there should be a continuous freezing transition between a weak-disorder phase, at large scales undistinguishable from the homopolymer
case,  and a strong-disorder glass phase with non-trivial scaling, and of
possible biological relevance since the conformation and properties of
RNA depends on the sequence disorder, i.e.\ on the primary structure.
This glass phase appears in the BH solution of (\ref{k1}) for the
$n=2$ replica case (instead of $n=0$ relevant for the disordered
case) and in numerical studies at strong disorder
\cite{BH,Krzakala02}.
Although the initial BH model is quite simple, this strong disorder
phase of random RNA appears to be highly non-trivial and quite
difficult to study, making it a challenging problem.

In \cite{LW} L\"assig and Wiese (LW) pioneered a field theoretical
(FT) approach for the transition to the glass phase. They showed their
model to be renormalizable at first order in perturbation theory and
calculated the critical exponents. Using a locking argument (see
below), the scaling exponents for random RNA in the {\em strong
disorder} phase were derived, in good agreement with
numerics \cite{BH,Krzakala02}.

It is important to understand if this approach defines a consistent
theory to all orders, and if the estimates of \cite{LW} for the
scaling exponents are reliable. We achieve this goal here.
Using a formulation of the LW model in terms of interacting random walks (RW) in $d=3$ dimensions, and FT tools developped for polymers and membranes
\cite{DDG3,WieseHabil} we show that the LW model  is
renormalizable to all orders.  Our formulation is more convenient for
calculations and allows us to derive new scaling relations between
exponents, and to calculate critical exponents at second order.

{T}{he} FT model of LW is defined through perturbation theory in the
disorder strength $g=\sigma^2$, the replica trick, and the continuum
limit where $L\to\infty$.  One introduces $n$ replicas of the system,
labelled by $\alpha=1,\dots ,n$.  For the free model (no disorder,
i.e. $\sigma=0$) the replicas are not coupled and the expectation
value of a product of $N$ contact operators $\Phi_\alpha(s_i,t_i)$ can
be computed exactly. It describes the constrained configuration with
$N$ fixed pairings $(s_i,t_i)$ $(i=1,\cdots,N)$, i.e.\ $N+1$ subrings of
backbone length $\ell_0,\cdots,\ell_N$ (with
$L=\ell_0+\cdots+\ell_N$).  As discussed above, this e.v.~is
\begin{equation}
\label{prodPhiev}
{\langle\Phi_\alpha(s_1,t_1)\cdots\Phi_\alpha(s_N,t_N)\rangle}_0=\ell_0^{-\rho_0}\ell_1^{-\rho_0}\cdots\ell_N^{-\rho_0}
\end{equation}
with $\rho_0=3/2$ if the $(s,t)$'s form a planar pairing, and $0$
otherwise. 
The partition function for a single free RNA strand then is
$Z_0^{(n=1)}={\langle 1\rangle}_0={L}^{-\rho_0}$.

The average over the disorder $\eta$ generates an attractive
interaction between two replicas, 
\begin{equation}
\label{HintPsi}
\mathcal{H}_i=
-g_0\sum_{\alpha<\beta}
\iint_{1\le s<t\le L}
\Psi_{\alpha\beta}(s,t)
\end{equation}
with coupling $g_0\propto\sigma^2$
and the overlap operator
\begin{equation}
\label{Psidef}
\Psi_{\alpha\beta}(s,t)=\Phi_{\alpha}(s,t)\Phi_{\beta}(s,t)\ .
\end{equation}
The quenched disorder average is obtained for $n\to 0$. The averaged
free energy $\overline F$ for a single strand is
\begin{equation}
\label{k3} \overline{F}=-\overline{\log Z_\eta}=\lim_{n\to
0}-{\partial\over\partial n}{Z}_{n}\ , \qquad
{Z}_{n}={\langle\mathrm{e}^{-\mathcal{H}_i}\rangle}_0\ .
\end{equation}
${Z}_{n}$ is the partition function for $n$ interacting
replicas. Similarly, the average of an observable $A$, $\overline
A$, is the $n\to 0$ limit of the average in the interacting theory.
{T}{hese} observables are calculated as perturbative series in the
disorder strength  $g_0$. They suffer from short-distance UV
divergences. Taking $\rho_0$ as an analytic regularization
parameter, $\rho_0\le 3/2$, LW show at first order that these UV
divergences are poles in $\epsilon=2\rho_0-2$ at $\epsilon=0$ and that
the theory is 1-loop renormalizable at $\epsilon=0$ ($\rho_0=1$).  An
UV-finite renormalized theory is defined through a renormalization of
the disorder strength $g_0$ and of the backbone length $L$.  This
allows to compute at first order the 
RG
$\beta$ function
for the disorder strength $g$.  It is found to have a UV fixed point
$g^*\,{>}\,0$ for $\epsilon \,{>}\,0$, in particular for the physical
case $\epsilon=1$,  $n=0$.  $g^*$ corresponds to the RNA
freezing transition. LW compute also the scaling dimensions
$\rho^\star$ and $\theta^\star$ of the operators $\Phi$ and $\Psi$ at
the transition.

Our goal is to construct a FT  which reproduces
(\ref{prodPhiev}).  For this we note that $\ell^{-\rho_0}$ is 
 the return probability at proper time $\ell$ for a free random
walk (RW) in $\mathbb{R}^d$ with $d=2\rho_0$.  Thus we introduce $n$
independent RW's,
$\mathbf{r}_\alpha,\,\alpha=1,\dots ,n$
($\mathbf{r}_\alpha(s)=\{r_\alpha^\mu(s);\,\mu=1,\dots ,d\}$). In order to
keep only planar pairings we use  $n\times N$ pairs of auxiliary
fields $\gamma_a^\alpha(s)$ and $\tilde\gamma_a^\alpha(s)$
($a=1,\ldots, N$).  The free
model is given by the action ($\,\dot{\ }:=\partial/\partial s$)
\begin{equation}
\label{k4} 
\mathcal{S}_0=\sum_{\alpha=1}^n\int_0^L \!\rmd s\,{1\over
4}\Big[\dot{ \mathbf{r}}_\alpha(s)\Big]^2 
\,+\,\sum_{\alpha=1}^n\sum_{a=1}^N\int_0^L
\!\rmd s\,\tilde\gamma_a^\alpha(s)\dot{\gamma}_a^\alpha(s)
\end{equation}
The propagators for $r$ and $\gamma,\tilde\gamma$ are
\begin{equation}
\label{k5} 
\frac{1}{2}\left< \left[ r_\alpha^\mu(s)
-r_\beta^\nu(t)\right]^{2}\right>_0=\delta_{\alpha\beta}\delta^{\mu\nu}|s-t|
\end{equation}
\begin{equation}
{\langle \tilde{\gamma}^\alpha_a(s)
\gamma_b^\beta(t)\rangle}_0=\delta^{\alpha\beta}\delta_{ab}\theta(t-s)
\ ,
\quad\langle\gamma\gamma\rangle=\langle\tilde\gamma\tilde\gamma\rangle=0
\nonumber 
\ ,
\end{equation}
where $\theta(s)=1$ if $s>0$, and $0$  otherwise.

{T}{he} key point is that in the large-$N$ limit, the
observables for a strand of length $L$  in the LW model correspond to the
partition function for a closed RW in our model, with specific
boundary conditions (the end points are fixed and there is a creation
operator $\tilde\gamma$ at the origin and an annihilation operator
$\gamma$ at the end).  
The contact operator $\Phi$ changes to
\begin{equation*}
\label{PhiRW}
\Phi_{\!\alpha}(s,t){=}{1\over N}\!\sum_{a,b}\!\gamma_a^\alpha\!(s)\tilde{\gamma}_b^\alpha\!(s)\delta^d[\mathbf{r}_\alpha(s){-}\mathbf{r}_\alpha(t)]\gamma_b^\alpha\!(t)\tilde{\gamma}_a^\alpha\!(t).
\end{equation*}
{T}{he} pair-contact operator $\Psi_{\alpha \beta} (s,t)$ is
still given by (\ref{Psidef}), and the interaction by
(\ref{HintPsi}). The auxiliary fields $\gamma$ and $\tilde \gamma$
allow to select planar diagrams by taking $N\to \infty$. For the
analysis of the UV-divergences, they are mere spectators. Their
importance is that they allow to write an action, and thus to apply
established tools to prove renormalizability, and to obtain exponents
at higher orders. For the sake of simplicity, we shall not write the
$\gamma$'s explicitly in the following.
We also note that a $1/N$-expansion is feasible, similar in spirit to \cite{OrlandZee2002} for the homopolymer problem.

T{he} model defined by (\ref{k4}) belongs to a class of theories with
multilocal interactions, including the Edwards model of polymers and
Self Avoiding Manifolds (SAM) \cite{DDG3,WieseHabil}.  Its
short-distance singularities can be studied by the same Multilocal
Operator Product Expansion (MOPE).  Indeed, the operator $\Psi$ is a
product of bi-local operators $\delta[\mathbf{r}_{\alpha } (s)
-\mathbf{r}_{\alpha } (t)]$ and of auxiliary fields
$\gamma\tilde\gamma$.  These auxiliary fields have a very simple
propagator
and a trivial short-distance expansion, which is a product of $\theta$
functions, multiplying the MOPE of the $\delta$'s.  Let us give as
examples the configurations which encode the UV singularities relevant
at one loop.  The short-distance behavior of a single $\Psi$ is given
by
\begin{eqnarray}\label{Psi2One}
 \Psi_{\alpha\beta}(u,v) &\mbox{$\displaystyle \mathop{\simeq}_{v\to
u}$}&
|u-v|^{-d}\mathbf{1} \\
&&+|u-v|^{1-d}\Big[{\dot{\mathbf{r}}_\alpha(u)}^2\!\!+{\dot{\mathbf{r}}_\beta(u)}^2\Big]+\cdots \nonumber
\end{eqnarray}
and is depicted graphically as
\begin{equation*}
\parbox{.7in}{\includegraphics[width=.7in]{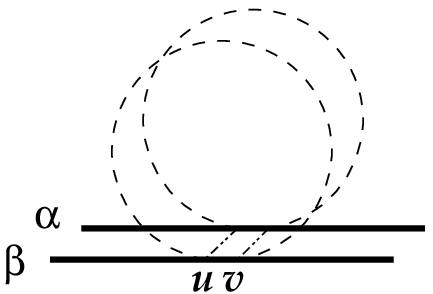}}\ \to\ 
\parbox{.7in}{\includegraphics[width=.7in]{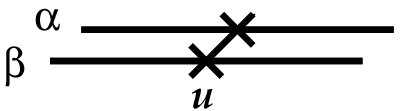}}
\end{equation*}
Similarly, two $\Psi$'s can coalesce into a single $\Psi$ in three
ways. Firstly as
\begin{equation}
\label{PsiPsi2Psi-a}
\Psi_{\alpha\beta}(u,\!v)\Psi_{\alpha\beta}(u',\!v')
\hskip-1.4em
\mathop{\simeq}_
{u'\to u,v'\to v}
\hskip-1.4em C(u'\!,\!u;v'\!,\!v)\Psi_{\alpha\beta}(u,v)
+\cdots
\end{equation}
that we depict as
\begin{equation*}
\parbox{1in}{\includegraphics[width=1.in]{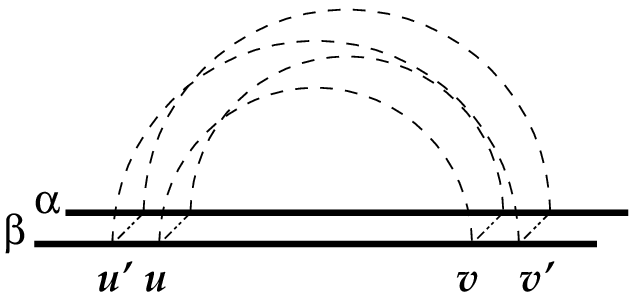}}\ \to\ 
\parbox{1in}{\includegraphics[width=1.in]{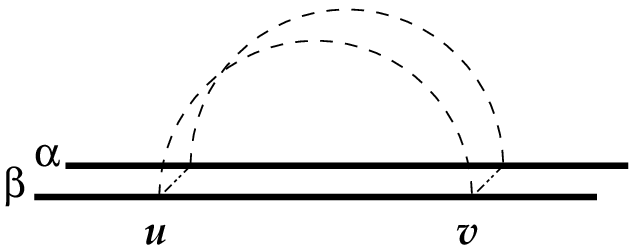}}
\end{equation*}
with $C(u',u;v',v)=\left(|u'-u|+|v'-v|\right)^{-d}$ the corresponding
MOPE coefficient.  Secondly as
\begin{equation}
\label{PsiPsi2Psi-b}
\Psi_{\alpha\beta}(u,\!u')\Psi_{\alpha\beta}(v,\!v')
\hskip-1.4em
\mathop{\simeq}_
{u\to v,u'\to v}
\hskip-1.4em D(u,\!u'\!,\!v)\Psi_{\alpha\beta}(v,\!v')
+\cdots
\end{equation}
with $D(u,\!u'\!,\!v)\!=\!|u'{-}u|^{-d}$ if $u{<}u'{<}v$ or
$v{<}u{<}u'$, and $=0$ otherwise, that we depict as
\begin{equation*}
\parbox{1in}{\includegraphics[width=1.0in]{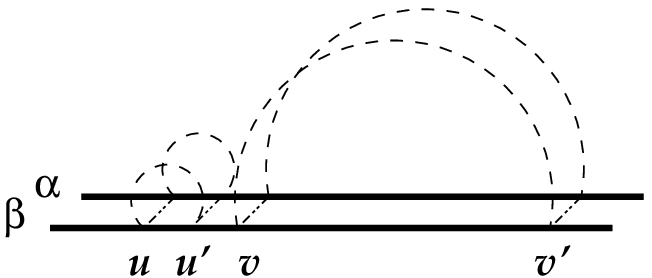}}+
\parbox{1in}{\includegraphics[width=1.0in]{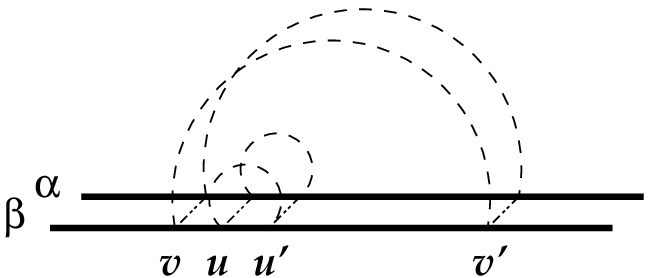}}
\to
\parbox{1in}{\includegraphics[width=1.0in]{Psi}}
\end{equation*}
Thirdly as
\begin{equation}
\label{PsiPsi2Psi-c}
\Psi_{\beta\gamma}(u,\!u')\Psi_{\alpha\beta}(v,\!v')
\hskip-1.4em
\mathop{\simeq}_
{u\to v,u'\to v}
\hskip-1.4em E(u,\!u'\!,\!v)\Psi_{\alpha\beta}(v,\!v')
+\cdots
\end{equation}
with $E(u,\!u'\!,\!v)\!=\!D(u,\!u'\!,\!v)$, that we depict as
\begin{equation*}
\parbox{1in}{\includegraphics[width=1.in]{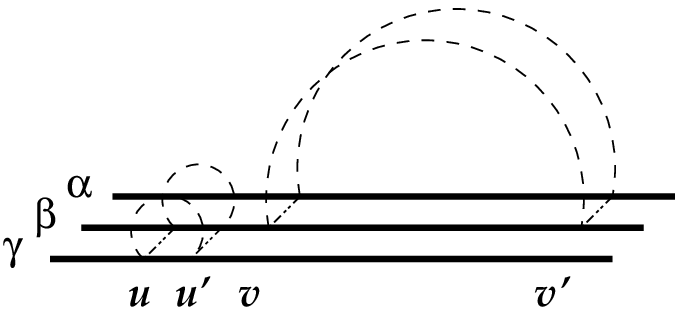}}+
\parbox{1in}{\includegraphics[width=1.in]{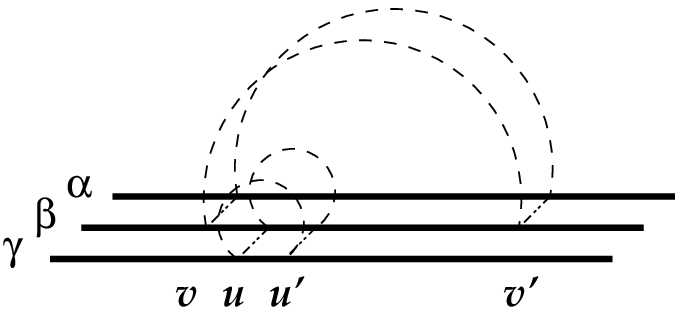}}\to
\parbox{1in}{\includegraphics[width=1.in]{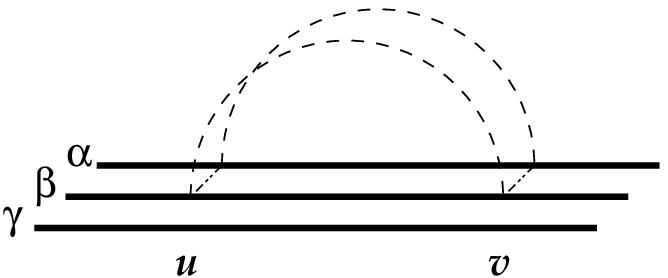}}
\end{equation*}
The perturbative expansion involves expectation values of integrals of
products of $\Psi$ operators.  The short-distance contribution
$u{\to}u'$ for a single $\Psi(u,u')$ in these integrals is given by
(\ref{Psi2One}) and produces an UV divergence.  The first term in
(\ref{Psi2One}) gives a UV pole at $d=1$ proportional to the insertion
of the unit operator $\mathbf{1}$, while the second one gives a pole
at $d=2$ proportional to the operator ${\dot{\mathbf{r}}}^{\,2}$.
Similarly, considering now the integrals involving two $\Psi$
operators, the integrals over $u$ and $u'$ ($v$ and $v'$ fixed) in
(\ref{PsiPsi2Psi-a},\ref{PsiPsi2Psi-b},\ref{PsiPsi2Psi-c}) give UV
poles at $d=2$, proportional to the operator $\Psi$.  In both cases,
the subdominant terms (represented by the $\cdots$) in the MOPE
involve higher dimensional multilocal operators, but do not give any
UV pole at $d\le 2$.  Note that although the l.h.s of
(\ref{PsiPsi2Psi-c}) involves three replicas, the dominant term on the
r.h.s. involves only the overlap operator between two replicas.
Also note that (\ref{PsiPsi2Psi-b}) does not contribute for polymers or SAM's,
since there is a third non-planar diagram, and the sum of all cancel.

This
analysis of the UV divergences through the MOPE at first order gives
the same results as in LW. 
It shows that our model is renormalizable to one loop at $d=2$, as
expected from the existence of an action, and dimensional analysis.

{O}{ur} formulation and the MOPE allow to extend this
analysis to all orders of perturbation theory \cite{DWlong}. The dimension $d=2\rho_0$
of imbedding space is a dimensional regularization parameter, and
short-distance UV divergences appear always as poles in
$\epsilon\!=\!d\!-\!2\!=\!2\rho_0\!-\!2$.
We have shown that the theory is UV finite for $\epsilon<0$ (apart
from a trivial ``vacuum energy divergence" proportional to the unity
operator $\mathbf{1}$).  For $\epsilon=0$ the only UV divergences are
proportional to the local operator $\dot{\mathbf{r}}_\alpha^{\,2}$ and
to the bilocal operator $\Psi_{\alpha\beta}$ ($\alpha\neq\beta$).  The
MOPE also generates multilocal operators involving more than 2
replicas, for instance the 3-replica operator
$\Pi_{\alpha\beta\gamma}=\Phi_\alpha\Phi_\beta\Phi_\gamma$.  However
these operators are not associated to UV divergences, and correspond
to irrelevant couplings.  The crucial point in this analysis is that,
since the interaction in the model involves 2 different replicas
$\alpha\neq\beta$, no UV divergence appears which is proportional to
the single-replica operator $\Phi_\alpha(u,v)$, although $\Phi$ is a
``dangerously" marginal operator (it is marginal at $\epsilon\!=\!0$
and relevant as soon as $\epsilon\!>\!0$, like $\Psi$).

{T}{he} renormalized UV finite theory is defined through the renormalized
action $\mathcal{S}_R$
\begin{equation}
\label{k10}
 \mathcal{S}_R=\sum_\alpha\int_0^L\!\!\rmd s\,{\mathbb{Z}\over
4}{\dot{\mathbf{r}}_\alpha^{\,2}}+g_R\mu^{-\epsilon}\mathbb{Z}_g\sum_{\alpha<\beta}\iint_{0\le
u<v\le L}\hskip-3em\Psi_{\alpha\beta}(u,v)
\end{equation}
$\mathbb{Z}$ and $\mathbb{Z}_g$ are the wave-function and
coupling-constant counterterms, and are series in $g_R$ whose
coefficients contain the poles in $1/\epsilon$. $\mu$ is the
renormalization mass scale.
The renormalized field $\mathbf{r}$ and coupling $g_R$ are related to
the bare ones $\mathbf{r}_B$ and $g_B$ by 
$\mathbf{r}_B=\mathbb{Z}^{1/2}\mathbf{r}$ and
$g_B=g_R\mathbb{Z}_g\mathbb{Z}^{d}\mu^{-\epsilon}$ (this differs from
the LW scheme where $s$ is renormalized instead of $\mathbf{r}$ and
where $g'_B=g_R\mathbb{Z}_g\mathbb{Z}^{2}\mu^{-\epsilon}$).  The RG
$\beta$ function for the coupling $\beta_g$ and the scaling dimension
$\chi_{r}$ for the field $\mathbf{r}$ in length-units are
\begin{equation}
\label{k11}
   \beta_g = -\mu \left.{\rmd g_R\over \rmd \mu}\right|_{g_B}\ ,\quad
   \chi_{r}={1\over 2}\left(1+\beta_g{d\log\mathbb{Z}\over
   dg}\right) .
\end{equation}
Since we {\em proved} renormalizability, and identified all
possible counterterms, we can simplify calculations by using open RWs
instead of closed ones, eliminating the $\delta$-function for the
closure. Although not all correlation functions are directly
interpretable in terms of RNA strands, they are renormalized by the
same counterterms, except for one additional boundary term
$\mathbb{Z}_1$ for each end of the RW,
$S_R^{\mathrm{open}}=S_R^{\mathrm{closed}}+2\mathbb{Z}_1$.
The (Fourier transformed) partition function $Z^{(1)}$  for a \emph{single} free open RW then is
\begin{equation}
\label{Zopen0}
\parbox{.4in}{\includegraphics[width=.4in]{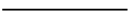}}=
Z^{(1)}(\vec q)=
\Big\langle{\prod_{\alpha=1}^{n}\mathrm{e}^{-\vec q(\vec r_\alpha({\scriptscriptstyle{L}})-\vec r({\scriptscriptstyle{0}}))}}\Big\rangle
=
\mathrm{e}^{-n\vec q\,^2L}
\ .
\end{equation}
The first-order correction to (\ref{Zopen0}) in the disorder strength
is given by the following diagram
\begin{equation}
\label{k12} 
\parbox{.8in}{\includegraphics[width=.8in]{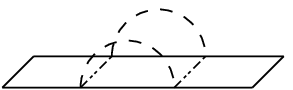}}=g_0{n(n-1)\over
2}\iint_{0<u<v<L}\hskip-3em\mathrm{e}^{\vec
q\,^2(2|v-u|-nL)}|v-u|^{-d}
.
\end{equation}
The last integral is UV divergent when $u\,{\to}\, v$ (bulk) and $ u,v\,{\to}\, 0$
or $u,v\,{\to}\, L$ (boundary). The corresponding UV pole in $\epsilon=0$
has residue $(1-2L\vec q\,^2)\exp(-n\vec q\,^2L)$.  This 
partition function for a single open RW is renormalized as
\begin{equation}
\label{k13}
Z^{(1)}_R(\vec q,g_R)=
Z_B^{(1)}(\mathbb{Z}^{-1/2}\vec q,g_B)\,\mathrm{e}^{-2\mathbb{Z}_1}
\ .
\end{equation}
This implies that in the MS scheme the counterterms are at first order
$\mathbb{Z}=1+g_R(n-1)/\epsilon$, $\mathbb{Z}_1=g_R(n-1)/4\epsilon$.

{T}{o} compute $\mathbb{Z}_g$ it is simpler to consider the (connected)
partition function $Z^{(2)}$ for \emph{two distinct} interacting open RWs. At first
order in $g_0$ it is given by 
\begin{equation}
\label{k14}
\parbox{.7in}{\includegraphics[width=.7in]{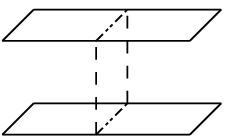}}=Z^{(2)}=g_0{n(n-1)\over 2}L^2\ .
\end{equation}
At order $g_0^2$ there are 4 UV divergent diagrams, with MOPE given in
(\ref{PsiPsi2Psi-a}), (\ref{PsiPsi2Psi-b}), and (\ref{PsiPsi2Psi-c}), 
\begin{center}
\includegraphics[width=.82in]{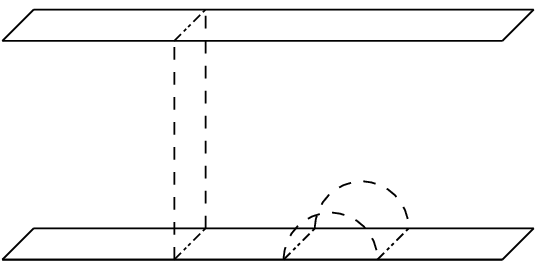}
\includegraphics[width=.82in]{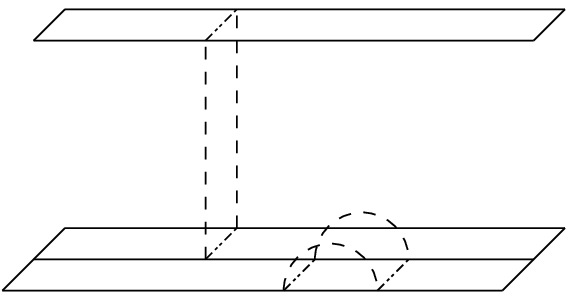}
\includegraphics[width=.82in]{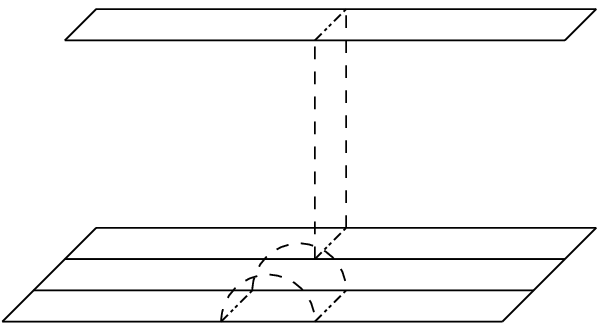}
\includegraphics[width=.82in]{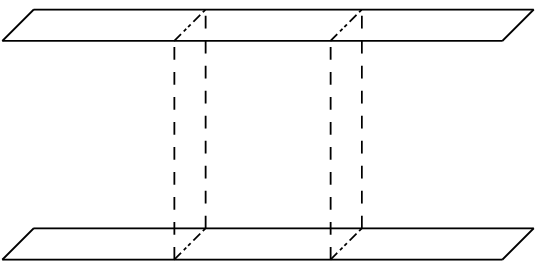}
\end{center}\par\noindent
and the corresponding function is renormalized as
\begin{equation}
\label{k15}
Z^{(2)}_R(g_R)=
\mathbb{Z}^{-d}
Z_B^{(2)}(g_B)\mathrm{e}^{-4\mathbb{Z}_1}\ .
\end{equation}
Care was taken to account for the (missing) zero-mode due to the
$\delta$-distribution between the 2 replicas, resulting in the factor
of $\mathbb{Z}^{-d}$. 
To subtract the UV pole at $\epsilon=0$ the counterterm is
$\mathbb{Z}_g=1+g_R(7-4n)/\epsilon$.

\begin{figure}[t]
\begin{center}
\includegraphics[width=2.5in]{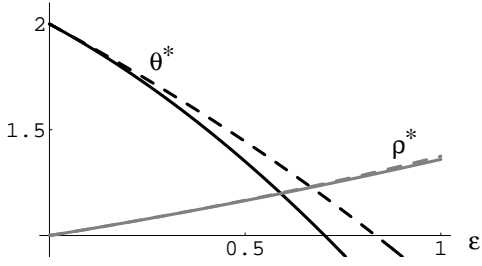} 
\caption{Results for
$\theta^{*}$ (black) and $\psi^{*}$ (grey) at 1- (dashed) and 2-loop
(solid) order.}  \label{k22}
\end{center}
\end{figure}

{T}{his} scheme can be continued to 2-loop order \cite{DWlong}. To simplify the
results we subtract minimally $\mathbb{Z}$ and $\mathbb{Z}_{g}\mathbb{Z}^{d}$. Also
$\rho (g_{r})$, the dimension of $\Phi$, can be calculated by
considering a 2-RW partition function $Z_{\Phi }$ with one $\Phi$
connecting the 2 RW's.
We obtain the RG functions at 2 loops:
\begin{eqnarray}\label{beta}
\beta_g(g_R) &=& -\epsilon g_R+(5-2n)g_R^2+(3-2n)(5-2n)g_R^3
\nonumber \\
{\chi}_{r}(g_R) &=& {1\over 2}-{n-1\over 2}g_R-{(n-1)(4-3n)\over
2}g_R^2
 \\
\rho (g_R) &=& 1+{\epsilon\over 2}+(n-1)g_R+{(n-1)(3-4n)\over
2}g_R^2\nonumber
\ .
\end{eqnarray}
At one loop our results agree with those of LW \cite{LW}, upon identifying
$\beta=\beta_{\scriptscriptstyle{\mathrm{LW}}}$, and
$\chi_r=1/ (2\gamma_{\scriptscriptstyle{\mathrm{LW}}})$. 
{I}{n} the physical case of {\em random} RNA ($n=0$), our
2-loop results confirm the existence of a UV fixed point (FP) (in our
scheme at $g^*={1\over 5}\epsilon-{3\over 25}\epsilon^2$),
describing the freezing transition.
The anomalous dimensions of $\Psi$ and $\Phi$ at this FP are
\begin{align}
\label{k20}
 \theta^*   
&=\Delta_\psi(g^*)=2+\beta_g'(g^*)=2-\epsilon-{3\over 5}\epsilon^2   \\
  \rho^* & =\Delta_\Phi(g^*)=1+{3\over 10}\epsilon+{3\over 50}\epsilon^2\ .
\end{align}
On Fig.~\ref{k22} we plot $\theta^*$ (black) and $\rho^*$ (grey)
from $\epsilon=0$ to $\epsilon=1$ (physical case). The full curves are
our 2-loop results, the dashed ones the 1-loop estimates of \cite{LW}.
The 2-loop corrections do not change much the estimate for $\rho^*$,
but do change it quantitatively for $\theta^*$. This defines another
dependent exponent, the roughness $\zeta$ of the height-field $h
(r):=\sum_{s<r}\sum_{{t>r}}\Phi (s,t)\sim r^{\zeta}$, as $\zeta
=2-\rho$ \cite{LW}. 
{S}{ince} $\Phi_{\alpha} (s,t)\le 1$, $\left< \Psi_{\alpha
\beta}\right> (s,t)\le \left<\Phi_{\alpha} (s,t) \right>$ and thus
$\rho\le\theta$. This bound is clearly violated by our results for
$\epsilon>\epsilon_\mathrm{c}\simeq0.59$. According to LW in this
regime the two replicas are locked into a single conformation,
$\rho^*=\theta^*$ and the exponents in the glass phase equal those at
the transition
\begin{equation}
\label{o1}
\rho_{\mathrm{glass}}=\rho^*|_{\epsilon=1}\simeq 1.36, \quad \zeta_{\mathrm{glass}}=2{-}\rho^*\simeq 0.64\ .
\end{equation}
Finding a small $\epsilon^2$ correction to $\rho^*$ is important,
since it validates the estimates of \cite{LW} for the exponents of
random RNA. Numerical results obtained by Krzakala et
al.~\cite{Krzakala02} in agreement with Bundschuh et
al.~\cite{BH,Bundschuh.pc} give
\begin{equation}\label{o2}
\rho_{\mathrm{glass}}\simeq 1.34\pm 0.003,\quad \zeta_{\mathrm{
glass}}\simeq0.67 \pm 0.02
\end{equation}
{I}{n} \cite{LW} it was also conjectured that the 
dimensions of $\Phi$ and $\mathbf{r}$ are not independent.  Our
formalism shows that this is correct and gives an exact relation
between $\chi_r$ and $\rho$. We remark that the partition function
$Z_{\Phi}$ for one $\Phi$ connecting two strands is equivalent to that
of two single strands, upon marking a single point on each strand,
i.e.\
\begin{equation}
\label{o3}
Z_{\Phi} =\displaystyle   \frac{1}{n^{2}} \Big[-\frac{\partial}{\partial
{q^{2}}}Z^{(1)} (q) \Big|_{q=0} \Big]^2
\ .
\end{equation}\hskip0mm

\noindent At $g^*$, together with
$\zeta + \rho =2$ \cite{LW}, this gives 
\begin{equation}\label{o5}
\zeta^{*}  = 2-\rho^*=(2-\epsilon) \chi_{r}^{*}\ .
\end{equation}
{I}{n} conclusion, our results for the RNA freezing transition are as
follows: First we give a new field theoretical formulation of
L\"assig-Wiese \cite{LW}, and prove that their model is renormalizable
to all orders in perturbation theory.  As a consequence we show that
the $\epsilon$-expansion scheme of \cite{LW} is well defined.  Second
our formulation allows to simplify the perturbative calculations, in
particular by considering open interacting RW's instead of closed RNA
strands. We perform the first 2-loop calculation for the critical
exponents $\theta$ and $\rho$, and show that it does not much correct
the LW estimate for $\rho$.  Third, we derive a new scaling law
relating the dimensions of the height field $h$ and $r$. Finally let
us mention that we have applied our methods to the denaturation of
random RNA under tension, which allows to calculate the extension
force curve \cite{DHW}.

K.W.~thanks M.~L\"assig for the stimulating collaboration which raised
many of the questions addressed above.

\vfill

\begin{thebibliography}{99}
\bibitem{seq}
For sequence disorder with $b=4$ different bases, the pair energy
variables $\eta(s,t)$ are correlated. However, treating them as
independent random variables is numerically an excellent
approximation~\cite{BH,Krzakala02}.
\end{thebibliography}
\end{document}